\documentclass[letterpaper,english,aps, reprint, superscriptaddress, showkeys]{revtex4-1}
\usepackage[T1]{fontenc}
\usepackage[latin9]{inputenc}
\usepackage{amsmath}
\usepackage{amssymb}
\usepackage{graphicx}
\usepackage{microtype}

\makeatletter

\pdfpageheight\paperheight
\pdfpagewidth\paperwidth

\providecommand{\tabularnewline}{\\}

\usepackage{hyperref}
\usepackage{amsmath,calc}
\usepackage{array}
\usepackage{booktabs}
\usepackage[T1]{fontenc}

\newcolumntype{C}[1]{>{\centering\arraybackslash$}m{#1}<{$}}
\newlength{\mycolwd}                                         
\makeatother

\usepackage{babel}
\begin{document}
\title{Spin Dynamics Investigation of Quasi-Frozen Spin Lattice for EDM Searches}
\author{\firstname{Eremey} \surname{Valetov}}
\email{valetove@msu.edu}

\affiliation{IKP, Forschungszentrum Jülich, Germany}
\affiliation{Michigan State University, East Lansing, MI 48824, USA}
\author{\firstname{Yurij} \surname{Senichev}}
\affiliation{IKP, Forschungszentrum Jülich, Germany}
\author{\firstname{Martin} \surname{Berz}}
\affiliation{Michigan State University, East Lansing, MI 48824, USA}
\collaboration{On behalf of the JEDI Collaboration}
\date{September 27, 2016}
\begin{abstract}
The Quasi-Frozen Spin (QFS) method was proposed by Yu. Senichev\emph{
et al.} in \cite{Senichev2015} as an alternative to the Frozen Spin
(FS) method \cite{pEDM} for the search of deuteron electric dipole
moment (dEDM). The QFS approach simplifies the design of the lattice.
In particular, small changes to the currently operating COSY storage
ring will satisfy the QFS condition. Spin decoherence and systematic
errors fundamentally limit EDM signal detection and measurement. Our
QFS implementation method includes measurement of spin precession
in (1) the horizontal plane to calibrate the magnetic field when changing
field polarity and (2) the vertical plane to search for EDM. To address
systematic errors due to element misalignments, we track particle
bunches in forward and reverse directions. We modeled and tracked
two QFS and one FS lattice using the code \emph{COSY INFINITY}. The models include
normally distributed random variate spin kicks in magnetic dipoles
and combined electrostatic and magnetic field elements. We used \emph{Wolfram
Mathematica} programs to partially automate lattice input file generation
and tracking output data analysis. We observed indications that the
QFS method is a viable alternative to the FS method.
\end{abstract}
\maketitle

\section{Introduction}

\subsection{The Quasi-Frozen Spin Concept}

The Frozen Spin (FS) concept for search of a deuteron electric dipole
moment (dEDM) was proposed in the BNL Report \cite{pEDM}. The idea
of the FS concept is that (1) the spin vector is aligned with the
particle momentum vector as the particle moves in a lattice and (2)
the radial electrostatic field results in torque on the spin vector,
rotating it out of the midplane.

The generalized Thomas-BMT equation is
\[
\frac{d\overrightarrow{S}}{dt}=\overrightarrow{S}\times\left(\vec{\Omega}_{\mathrm{MDM}}+\vec{\Omega}_{\mathrm{EDM}}\right)\mathrm{,}
\]
where the magnetic dipole moment (MDM) angular frequency is 
\[
\vec{\Omega}_{\mathrm{MDM}}=\frac{e}{m}\left[G\vec{B}-\left(G-\frac{1}{\gamma^{2}-1}\right)\frac{\vec{E}\times\vec{\beta}}{c}\right]
\]
and the electric dipole moment (EDM) angular frequency is
\[
\vec{\Omega}_{\mathrm{EDM}}=\frac{e}{m}\frac{\eta}{2}\left[\frac{\vec{E}}{c}+\vec{\beta}\times\vec{B}\right]\mathrm{.}
\]

The Quasi-Frozen Spin (QFS) concept is based on the FS concept, but
the requirement that spin needs to be aligned with momentum is relaxed:
in QFS, spin is aligned with momentum \textit{on average}. The QFS
condition is expressed as
\[
\gamma G\Phi_{B}=\left[\frac{1}{\gamma}\left(1-G\right)+\gamma G\right]\Phi_{E}\mathrm{,}
\]
where $\Phi_{B}$ and $\Phi_{E}$ are the average angles of momentum
rotation due to magnetic and electric fields, respectively.

In our spin decoherence and systematic errors studies concerning FS
and QFS concepts, we consider three lattices codenamed \emph{Senichev
6.3}, \emph{Senichev E+B}, and \emph{Senichev BNL}. Spin decoherence
in these lattices is suppressed by the following:
\begin{enumerate}
\item RF cavity: first and, partially, second order components by mixing
the particles relative to the average field strength, averaging out
the $\triangle\gamma G$ for each particle.
\item Sextupoles: the remaining second order component, which is due to
the average of $\triangle\gamma G$ being different for each particle.
\end{enumerate}

\subsection{Senichev FS and QFS Lattices}

\paragraph{Senichev 6.3 QFS lattice \cite{Senichev2015}}

This lattice consists of 4 straight, 4 magnetic, and 4 electrostatic
sections. It has a characteristic ``hourglass'' shape. A variation
of this lattice can be implemented with relatively minor changes to
a number of existing lattices, including the Cooler Synchrotron COSY
at Forschungszentrum Jülich.

\paragraph*{Senichev E+B QFS lattice \cite{Senichev2015a}}

This lattice consists of 2 straight, 4 magnetic, and 2 straight E+B
(that is, combined electrostatic and magnetic) sections. Straight
E+B static Wien filter elements are used instead of the curved electrostatic
deflectors (1) to remove nonlinear components due to curvature in
cylindrical electrostatic element electrodes and (2) to simplify the
system from the engineering perspective.

\paragraph*{Senichev BNL FS lattice}

This lattice consists of 2 straight and 2 curved E+B sections. The
design of this lattice implements the FS method and is similar to
the lattice described in the $\mathtt{RingLat}$ Appendix of \cite{pEDM}.
The curved E+B sections use the curved E+B element proposed in \cite{pEDM}.

The straight sections in the lattices provide for the accelerating
station, sextupoles, beam injection and extraction, and measurement
equipment, including the polarimeter.

\section{Computational Methods}

\subsection{Computational Software}

We use the code \emph{COSY INFINITY} \cite{COSY} for various spin
tracking calculations, including:
\begin{enumerate}
\item manual and automatic spin decoherence optimization by sextupole family
strengths;
\item investigation of spin decoherence growth as a function of the number
of turns; and
\item study of the effects of systematic errors on spin decoherence.
\end{enumerate}
\par

We note \emph{COSY INFINITY} uses the following beamline coordinate
system:\medskip{}

\begin{tabular}{cc}
$\begin{aligned}r_{1} & =x\mathrm{,}\\
r_{3} & =y\mathrm{,}\\
r_{5} & =l=-\left(t-t_{0}\right)v_{0}\frac{\gamma}{1+\gamma}\mathrm{,}\\
r_{7} & =\delta_{m}=\left(m-m_{0}\right)/m\mathrm{,}
\end{aligned}
$ & $\begin{aligned}r_{2} & =a=p_{x}/p_{0}\mathrm{,}\\
r_{4} & =b=p_{y}/p_{0}\mathrm{,}\\
r_{6} & =\delta_{K}=\frac{K-K_{0}}{K}\mathrm{,}\\
r_{8} & =\delta_{z}=\left(z-z_{0}\right)/z_{0}\mathrm{.}
\end{aligned}
$\tabularnewline
\end{tabular}\medskip{}
where $x$ and $y$ are local transversal spacial coordinates in meters;
$p$, $K$, $v$, $t$, $\gamma$, $m$, and $z$ are the momentum,
kinetic energy, velocity, time of flight, total energy over $mc^{2}$,
mass, and charge respectively; and the index $0$ refers to the reference
particle.

We use \emph{Wolfram Mathematica} 10.4 for:
\begin{enumerate}
\item automated preparation of \emph{COSY INFINITY} input files from templates
using markers and regular expressions; and
\item storage, processing, quality assurance, and report generation using
data from the \emph{COSY INFINITY} output files.
\end{enumerate}

\subsection{Spin Decoherence Optimization}

We optimize the spin decoherence by sextupole strengths as follows:
\begin{enumerate}
\item We manually minimize spin decoherence up to $\pm0.2\:\mathrm{T/m}$
by sextupole family strengths in the $x-a$, $y-b$, and $l-\delta_{K}$
planes with a set of RF cavity frequencies and voltages.
\item We completed the optimization automatically using the LMDIF optimizer
that is built into \emph{COSY INFINITY}. At optimal values, the sextupole
family strength typically has a $10^{-3}\:\mathrm{T/m}$ error without
a significant impact on the spin decoherence.
\end{enumerate}

\subsection{Reverse Spin Transfer Map}

Since we track spin motion in lattices in both forward and reverse
directions, we need to compute reverse orbital and spin transfer maps.
There is already a built-in procedure for computation of the reverse
orbital transfer map in \emph{COSY INFINITY}. We have introduced a
procedure to calculate the reverse spin transfer map in 2016.

Consider a spin transfer map $M:X_{i}\rightarrow X_{i+1}$, where
both $X_{i}$ and $X_{i+1}$ are the 3D sphere $S^{3}$. Taking into
account the nonlinear terms, $M$ is a $3\times3$ matrix with differential
algebra-valued elements, i.e. $M\in\mathrm{SO}_{3}\left(_{n}D_{v}\right)$
(whereas $M$ would be in $\mathrm{SO}_{3}\left(\mathbb{R}\right)$
in the linear case). The inverse spin transfer map is the inverse
matrix $M^{-1}:X_{i+1}\rightarrow X_{i}$.

The time reversal results in the sign change of momentum (coordinates
$a$ and $b$) and the longitudinal offset (coordinate $l$) \cite[p.147]{ModernMapMethods}.

To obtain the reverse spin transfer map $M^{\mathrm{R}}$, we apply
the reversion transformation to the inverse spin transfer map:
\[
M^{\mathrm{R}}=\hat{R}_{\mathrm{S}}\cdot(M^{-1}\circ\hat{R})\cdot\hat{R}_{\mathrm{S}}\mathrm{,}
\]
where \begin{equation}
\hat{R}=\left (
\begin{array}{*{9}{@{}C{\mycolwd}@{}}}
    1 & 0 & 0 & 0 & 0 & 0\\
    0 & 1 & 0 & 0 & 0 & 0\\
    0 & 0 & 1 & 0 & 0 & 0\\
    0 & 0 & 0 & -1 & 0 & 0\\
    0 & 0 & 0 & 0 & -1 & 0\\
    0 & 0 & 0 & 0 & 0 & -1\\
\end{array}
\right )
\end{equation}acts on the \emph{COSY INFINITY} 6-dimensional phase space coordinates
$\left(x,y,\delta_{K},a,b,l\right)$ and\begin{equation}
\hat{R}_{\mathrm{S}}=\left (
\begin{array}{*{9}{@{}C{\mycolwd}@{}}}
    1 & 0 & 0 \\
    0 & 1 & 0 \\
    0 & 0 & -1 \\
\end{array}
\right )
\end{equation}acts on the spin vector coordinates $\left(s_{x},s_{y},s_{z}\right)$.

\subsection{Error Field Implementation}

According to the Thomas-BMT equation, a small perturbation of the
magnetic field acts, to the first order, as a small rotation on the
spin vector. We have implemented field errors as small, normally distributed
spin kicks applied to the magnetic dipoles or combined E+B elements.
The spin elements are interposed automatically into the \emph{COSY
INFINITY} code using one of the \emph{Mathematica} notebooks.

\section{Spin Decoherence Study}

In 2015, we studied the spin decoherence in the three \emph{Senichev}
lattices \cite{Senichev2015a,Valetov2015}. We presented a summary
of this study at SPIN 2016 for reference and comprehensiveness purposes.

The primary findings were:
\begin{enumerate}
\item With an optimized sextupole family strength, the spin decoherence
often remains in the same range for at least $5\times10^{5}$ turns.
This is promising in respect to the requirement of maintaining a spin
coherence time of no less than $\sim1000$ seconds to possibly build
a measurable EDM signal.
\item The QFS structure decoherence is at least as good as, or better than,
the FS structure decoherence.
\end{enumerate}

\section{Systematic Errors Study}

Systematic errors due to imperfections in the physical facility can
create a fake EDM signal. Considering (1) the Thomas-BMT equation,
(2) the reversal of the magnetic field in the reverse lattice, and
(3) the lattice structure's imposition of an interdependence on the
strengths of magnetic bends and electrostatic deflectors, we focus
our attention on the rotational magnet misalignments.

We have studied the effect of rotational magnet misalignments on spin
dynamics, namely spin decoherence and frequencies of rotation in a
vertical plane, in QFS and FS structures. The magnetic error field
components $B_{x}$ and $B_{z}$ are the most relevant to the detection
of an EDM signal because the $B_{y}$ component only results in differential
rotation in the horizontal plane.

\subsection{Clockwise and Counterclockwise Lattice Traversal}

To extract the EDM signal, we propose to track polarized particle
bunches in the QFS and FS lattices in both clockwise (CW) and counterclockwise
(CCW) directions. We consider the CW direction to be forward and the
CCW direction to be reverse. We use the fact that, in the linear approximation,
the reverse spin transfer map coincides with the inverse spin transfer
map. The reversal of a transfer map reverses the direction of the
magnetic field but does not affect the electrostatic field.

\subsection{$B_{x}$ Error Field Component}

An approximate solution of the Thomas-BMT equation in a magnetic dipole
with an error field $B_{x}$ and initial conditions $\vec{S}=\vec{e}_{z}$
and $\Omega_{z}=0$ is
\begin{align*}
S_{x}\left(t\right) & =\frac{\Omega_{y}\sin\left(\sqrt{\Omega_{x}^{2}+\Omega_{y}^{2}}t\right)}{\sqrt{\Omega_{x}^{2}+\Omega_{y}^{2}}}\mathrm{,}\\
S_{y}\left(t\right) & =-\frac{\Omega_{x}\sin\left(\sqrt{\Omega_{x}^{2}+\Omega_{y}^{2}}t\right)}{\sqrt{\Omega_{x}^{2}+\Omega_{y}^{2}}}\mathrm{.}
\end{align*}

The rotation frequencies are $\varOmega_{x}^{\mathrm{CW}}=\varOmega_{B_{x}}^{\mathrm{CW}}+\varOmega_{\mathrm{EDM}}$
and $\varOmega_{x}^{\mathrm{CCW}}=-\varOmega_{B_{x}}^{\mathrm{CCW}}+\varOmega_{\mathrm{EDM}}$
in the vertical plane and $\varOmega_{y}=0+\left\langle \delta\varOmega_{\mathrm{decoh}}\right\rangle $
in the horizontal plane.

It is necessary to (1) minimize the decoherence in the vertical plane
$\sigma_{\varOmega_{B_{x}}}$ the same way as in the horizontal plane
using the RF cavity and sextupole families and (2) minimize $\left|\varOmega_{B_{x}}^{\mathrm{CW}}-\varOmega_{B_{x}}^{\mathrm{CCW}}\right|$
by calibrating the fields of the magnets using the spin precession
frequency in the horizontal plane.

Rotation frequency due to EDM is obtained by $\varOmega_{\mathrm{EDM}}=\left(\varOmega_{x}^{\mathrm{CW}}+\varOmega_{x}^{\mathrm{CCW}}\right)/2$.

\subsection{$B_{z}$ Error Field Component}

From an approximate solution of the Thomas-BMT equation in a magnetic
dipole with an error field $B_{z}$, $\Omega_{z}=\Omega_{B_{Z}}$,
$\Omega_{y}=\left\langle \varOmega_{\mathrm{decoh}}\right\rangle $,
$\Omega_{B_{z}}\ll\left\langle \varOmega_{\mathrm{decoh}}\right\rangle $,
and initial conditions $\vec{S}=\vec{e}_{z}$ and $\Omega_{x}=0$
\begin{align*}
\left\langle S_{x}\left(t\right)\right\rangle  & =\sin\left(\left\langle \varOmega_{\mathrm{decoh}}\right\rangle t\right)\mathrm{,}\\
\left\langle S_{y}\left(t\right)\right\rangle  & =\frac{\varOmega_{B_{Z}}}{\left\langle \varOmega_{\mathrm{decoh}}\right\rangle }\left[1-\cos\left(\left\langle \varOmega_{\mathrm{decoh}}\right\rangle t\right)\right]\mathrm{,}
\end{align*}
we see that the fake EDM signal has a factor of $\varOmega_{B_{z}}/\left\langle \delta\varOmega_{\mathrm{decoh}}\right\rangle $.

The method of error field component mitigation for $B_{x}$ is not
applicable to $B_{z}$.

We have to minimize $\varOmega_{B_{z}}$ to $\sim10^{-9}\:\mathrm{rad/s}$
using additional trim coils.

\subsection{Outcome of the $B_{x}$ and $B_{z}$ Component Mitigation Method}

With the application of the error component mitigation method outlined
here, a realistic estimate of measurement accuracy for $\mathrm{\varOmega_{EDM}}$
is $10^{-4}$ to $10^{-5}\:\mathrm{rad/s}$. As a result, the accuracy
of EDM signal measurement in one run is $10^{-24}$ to $10^{-25}\:e\mathrm{\cdot cm}$.
The accuracy of the EDM signal measurement after one year of measurement
may be $10^{-29}$ to $10^{-30}\:e\mathrm{\cdot cm}$.

\subsection{QFS/FS Conditions and Fringe Fields}

Considering the error field components, we need to study the resulting
spin dynamics and decoherence in the vertical plane for the \emph{E+B}
(QFS) and \emph{BNL} (FS) lattices. We have designed the \emph{E+B}
and \emph{BNL} lattices to satisfy the QFS or FS condition respectively
in the linear approximation. Taking into account the fringe fields,
the QFS/FS condition is approximately satisfied, resulting in spin
rotation in the horizontal plane. For the study of spin decoherence
in the horizontal plane, this is not an issue, because in that case
the spin decoherence is not significantly affected. However, when
error field components result in spin decoherence in the vertical
plane, the average spin direction in the horizontal plane affects,
among other things, the magnitude of the spin motive force in the
vertical plane. In the systematic errors study, we\emph{ }have to
consider in detail whether the QFS/FS condition is satisfied exactly
or approximately.

In a previous study, we attempted fitting the system configuration
to satisfy the nonlinear QFS/FS condition. This worked quite well
except that there was a small residual error in the meaasure of the
QFS/FS condition after the optimization. While orbital motion must
conserve energy and be symplectic, which is a nontrivial constraint,
the physical requirement for spin motion is only that the length of
the spin vector must be unity. We had recently introduced small spin
kicks after each turn (effectively an infinitely thin dipole acting
on the spin vector) to address the issue. In physical machine operations,
the field strengths can be adjusted to satisfy the QFS/FS condition
regarding the required precision.

\subsection{Spikes in Horizontal Spin Decoherence Measure at Zenith and Nadir}

With spin at the poles of the spherical coordinate system ($\theta=\pm\pi/2$),
we may observe apparent spikes in the horizontal spin decoherence.
This is not due to a physical effect, but rather due to defining the
measure of spin decoherence as $\sigma_{\phi}$ and $\sigma_{\theta}$.
To confirm this, we have analyzed a minimal test case with three particles
in \emph{Mathematica}. If we wished to avoid this effect, we could
use an alternative measure of spin decoherence, such as $\sigma_{S_{x}^{2}+S_{y}^{2}}$
in the horizontal plane and $\sigma_{S_{z}^{2}}$ in the vertical
plane. However, this alternative measure of spin decoherence has less
physical sense for large spin decoherence than with $\sigma_{\phi}$
and $\sigma_{\theta}$. The observed spikes in the horizontal spin
decoherence measure can be trivially accounted for in the analysis,
and their presence gives a useful visual indicator for spin motion
in the vertical plane.

\subsection{Spin-tracking Datasets}

We have produced more than 46 spin-tracking datasets in the systematic
errors study. This includes (1) \emph{Senichev 6.3}, \emph{E+B}, and
\emph{BNL} lattices; (2) optimization by SFP, SDP, SFP1, and SDP2
\footnote{In sextupole family names, 'F' denotes focusing, 'D' defocusing, 'P'
positive dispersion, and 'N' negative dispersion. The \emph{BNL} lattice
only has positive dispersion sextupole families.} sextupole families as well cases with no optimization; (3) $-0.5x$,
$0x$, $0.05x$, $0.5x$, $1x$, and $2x$ corrective spin kicks,
where $1x$ is defined to exactly satisfy the QFS/FS condition; and
(4) magnet misalignment angles of $10^{-4}\:\mathrm{rad}$ and, for
quality assurance purposes, $5\times10^{-5}\:\mathrm{rad}$.

Each dataset consists of 24 \emph{COSY} output files, including (1)
$B_{x}$ and $B_{z}$ error fields and no error field; (2) the CW
and CCW lattice traversal directions; (3) tracking in the horizontal
and the vertical plane, and (4) initial coordinates of the particle
bunch distributed in $x$ and $\delta_{K}$.

\subsection{Vertical Spin Decoherence, Approximate QFS/FS}

\begin{figure*}
\includegraphics[width=0.95\textwidth]{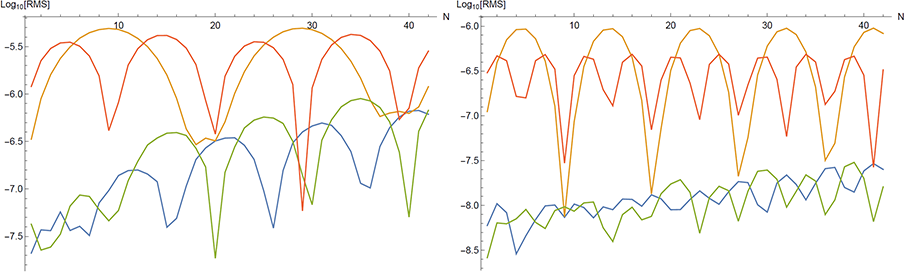}

\caption{Vertical spin decoherence (as $\log_{10}\left(\mathrm{rms}\right)$)
versus the number of turns (as $N=\mathrm{turns}/1000$), \emph{E+B}
lattice (left) and \emph{BNL} lattice (right), optimization by the
SDP sextupole family, and approximate QFS/FS (no corrective spin kick).\label{fig:1}}
\end{figure*}

See the plots in Fig. \ref{fig:1}
 for typical examples of vertical spin decoherence without corrective
spin kicks. With optimization by the SDP sextupoles, for the CW direction
with error field components $B_{x}$ (blue curve color) and $B_{z}$
(green), the vertical spin decoherence grows to $\sim10^{-6}\:\mathrm{rad}$
for the \emph{E+B} lattice and $\sim10^{-7.5}\:\mathrm{rad}$ for
the \emph{BNL} lattice in 420 thousand turns. With optimization by
the SDP sextupoles, for the CCW direction with error field components
$B_{x}$ (orange) and $B_{z}$ (red), the vertical spin decoherence
has the upper bound of $\sim10^{-5}\:\mathrm{rad}$ for the \emph{E+B}
lattice and $\sim10^{-6}\:\mathrm{rad}$ for the \emph{BNL} lattice
in 420 thousand turns.

\subsection{Vertical Spin Decoherence, Exact QFS/FS}

\begin{figure*}
\includegraphics[width=0.95\textwidth]{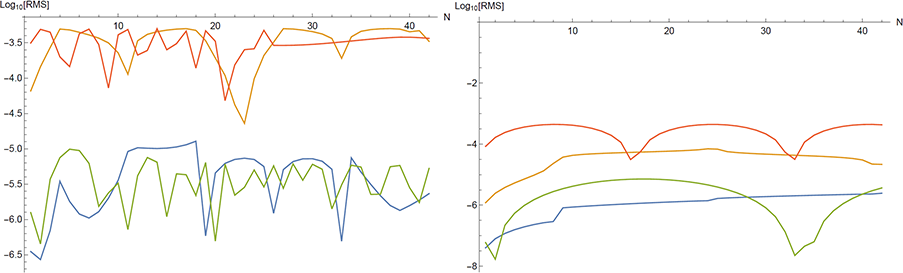}

\caption{Vertical spin decoherence (as $\log_{10}\left(\mathrm{rms}\right)$)
versus the number of turns (as $N=\mathrm{turns}/1000$), \emph{E+B}
lattice (left) and \emph{BNL} lattice (right), optimization by the
SDP sextupole family, exact QFS/FS ($1x$ corrective spin kick).\label{fig:2}}
\end{figure*}

See the plots in Fig. \ref{fig:2}
 for typical examples of vertical spin decoherence with $1x$ ccorrective
spin kicks. With optimization by the SDP sextupoles, for the CW direction
with error field components $B_{x}$ (blue curve color) and $B_{z}$
(green), the vertical spin decoherence has the upper bound of $\sim10^{-5}\:\mathrm{rad}$
for the \emph{E+B} lattice and $10^{-6.5}$ to $10^{-5}\:\mathrm{rad}$
for the \emph{BNL} lattice in 420 thousand turns. With optimization
by the SDP sextupoles, for the CCW direction with error field components
$B_{x}$ (orange) and $B_{z}$ (red), the vertical spin decoherence
has the upper bound of $\sim10^{-3}\:\mathrm{rad}$ for the \emph{E+B}
lattice and $10^{-4}$ to $10^{-3}\:\mathrm{rad}$ for the \emph{BNL}
lattice in 420 thousand turns.

\subsection{Summary of the Results}
\begin{enumerate}
\item The spin decoherence in the vertical plane is proportional to the
$B_{x}$ and $B_{z}$ error field components as expected.
\item When a spin kick is used for the exact QFS/FS, the vertical spin decoherence
in both lattices is about $10$ to $10^{2}$ higher, partly due to
the spin rotation in the horizontal plane that effectively acts as
an oscillation factor in the vertical spin motive force component
in case of inexact QFS/FS.
\item Because the sextupoles were optimized for the CW lattices, vertical
spin decoherence is somewhat higher for the CCW direction.
\item Some of the apparent periodic spikes in spin decoherence are due to
the use of the spherical coordinate system for the spin decoherence
measure.
\item With an optimized sextupole family strength and with exact QFS/FS
(via corrective spin kick), the vertical spin decoherence due to systematic
errors often remains in the same range for at least ${5\times10^{5}}$
turns in both the \emph{E+B} (QFS) and \emph{BNL} (FS) lattices. The
vertical spin decoherence in the \emph{E+B} (QFS) and \emph{BNL} (FS)
lattices is qualitatively very similar and quantitatively within about
1-2 orders.
\end{enumerate}

\section{Conclusion}
\begin{enumerate}
\item We estimate that in one year of measurement, the accuracy of EDM signal
measurement in QFS and FS lattices may be $10^{-29}$ to $10^{-30}\:e\mathrm{\cdot cm}$.
\item For at least ${5\times10^{5}}$ turns, the spin decoherence data in
the vertical plane due to error magnetic field components in the \emph{E+B}
(QFS) and \emph{BNL} (FS) lattices is

\begin{enumerate}
\item quantitatively within about 1\textendash 2 orders; and
\item qualitatively without significant differences.
\end{enumerate}
\item This systematic errors study is ongoing and will yield additional
results.
\item In the context of the systematic errors study, we will study the vertical
spin motion and spin decoherence with $B_{x}$ and $B_{z}$ error
field components while

\begin{enumerate}
\item slightly varying and optimizing the QFS/FS condition measure through
the electrostatic and magnetic field strengths; and
\item tracking the particle bunches for a larger number of turns.
\end{enumerate}
\end{enumerate}
\begin{acknowledgments}
This material is based upon work supported by the U.S. Department
of Energy, Office of Science, Office of High Energy Physics under
Award Number DE-FG02-08ER41546. This work has been supported by the
European Research Council under Grant Agreement 694340. This research
used resources of the National Energy Research Scientific Computing
Center, a DOE Office of Science User Facility supported by the Office
of Science of the U.S. Department of Energy under Contract No. DE-AC02-05CH11231.
\end{acknowledgments}

\bibliographystyle{apsrev4-1}
\nocite{*}
\bibliography{Proceedings}

\begin{thebibliography}{8}%
\makeatletter
\providecommand \@ifxundefined [1]{%
 \@ifx{#1\undefined}
}%
\providecommand \@ifnum [1]{%
 \ifnum #1\expandafter \@firstoftwo
 \else \expandafter \@secondoftwo
 \fi
}%
\providecommand \@ifx [1]{%
 \ifx #1\expandafter \@firstoftwo
 \else \expandafter \@secondoftwo
 \fi
}%
\providecommand \natexlab [1]{#1}%
\providecommand \enquote  [1]{``#1''}%
\providecommand \bibnamefont  [1]{#1}%
\providecommand \bibfnamefont [1]{#1}%
\providecommand \citenamefont [1]{#1}%
\providecommand \href@noop [0]{\@secondoftwo}%
\providecommand \href [0]{\begingroup \@sanitize@url \@href}%
\providecommand \@href[1]{\@@startlink{#1}\@@href}%
\providecommand \@@href[1]{\endgroup#1\@@endlink}%
\providecommand \@sanitize@url [0]{\catcode `\\12\catcode `\$12\catcode
  `\&12\catcode `\#12\catcode `\^12\catcode `\_12\catcode `\%12\relax}%
\providecommand \@@startlink[1]{}%
\providecommand \@@endlink[0]{}%
\providecommand \url  [0]{\begingroup\@sanitize@url \@url }%
\providecommand \@url [1]{\endgroup\@href {#1}{\urlprefix }}%
\providecommand \urlprefix  [0]{URL }%
\providecommand \Eprint [0]{\href }%
\providecommand \doibase [0]{http://dx.doi.org/}%
\providecommand \selectlanguage [0]{\@gobble}%
\providecommand \bibinfo  [0]{\@secondoftwo}%
\providecommand \bibfield  [0]{\@secondoftwo}%
\providecommand \translation [1]{[#1]}%
\providecommand \BibitemOpen [0]{}%
\providecommand \bibitemStop [0]{}%
\providecommand \bibitemNoStop [0]{.\EOS\space}%
\providecommand \EOS [0]{\spacefactor3000\relax}%
\providecommand \BibitemShut  [1]{\csname bibitem#1\endcsname}%
\let\auto@bib@innerbib\@empty
\bibitem [{\citenamefont {Senichev}\ \emph
  {et~al.}(2015{\natexlab{a}})\citenamefont {Senichev}, \citenamefont
  {Lehrach}, \citenamefont {Lorentz}, \citenamefont {Maier}, \citenamefont
  {Andrianov}, \citenamefont {Ivanov}, \citenamefont {Chekmenev}, \citenamefont
  {Berz},\ and\ \citenamefont {Valetov}}]{Senichev2015}%
  \BibitemOpen
  \bibfield  {author} {\bibinfo {author} {\bibfnamefont {Y.}~\bibnamefont
  {Senichev}}, \bibinfo {author} {\bibfnamefont {A.}~\bibnamefont {Lehrach}},
  \bibinfo {author} {\bibfnamefont {B.}~\bibnamefont {Lorentz}}, \bibinfo
  {author} {\bibfnamefont {R.}~\bibnamefont {Maier}}, \bibinfo {author}
  {\bibfnamefont {S.}~\bibnamefont {Andrianov}}, \bibinfo {author}
  {\bibfnamefont {A.}~\bibnamefont {Ivanov}}, \bibinfo {author} {\bibfnamefont
  {S.}~\bibnamefont {Chekmenev}}, \bibinfo {author} {\bibfnamefont
  {M.}~\bibnamefont {Berz}}, \ and\ \bibinfo {author} {\bibfnamefont
  {E.}~\bibnamefont {Valetov}} (\bibinfo {collaboration} {on behalf of the JEDI
  Collaboration}),\ }in\ \href
  {http://accelconf.web.cern.ch/AccelConf/IPAC2015/papers/mopwa044.pdf} {\emph
  {\bibinfo {booktitle} {Proceedings of IPAC 2015, Richmond, VA}}}\ (\bibinfo
  {year} {2015})\ \bibinfo {note} {{MOPWA044}}\BibitemShut {NoStop}%
\bibitem [{\citenamefont {Anastassopoulos}\ \emph {et~al.}(2008)\citenamefont
  {Anastassopoulos} \emph {et~al.}}]{pEDM}%
  \BibitemOpen
  \bibfield  {author} {\bibinfo {author} {\bibfnamefont {D.}~\bibnamefont
  {Anastassopoulos}} \emph {et~al.},\ }\href
  {https://www.bnl.gov/edm/files/pdf/deuteron_proposal_080423_final.pdf}
  {\enquote {\bibinfo {title} {{AGS Proposal: Search for a Permanent Electric
  Dipole Moment of the Deuteron Nucleus at the $10^{-29}\:e\cdot\mathrm{cm}$
  Level}},}\ }\bibinfo {howpublished} {{BNL Report, Brookhaven National
  Laboratory, Upton, NY}} (\bibinfo {year} {2008}),\ \bibinfo {note}
  {goo.gl/1K1TNq}\BibitemShut {NoStop}%
\bibitem [{\citenamefont {Senichev}\ \emph
  {et~al.}(2015{\natexlab{b}})\citenamefont {Senichev}, \citenamefont {Berz},
  \citenamefont {Valetov}, \citenamefont {Chekmenev}, \citenamefont
  {Andrianov},\ and\ \citenamefont {Ivanov}}]{Senichev2015a}%
  \BibitemOpen
  \bibfield  {author} {\bibinfo {author} {\bibfnamefont {Y.}~\bibnamefont
  {Senichev}}, \bibinfo {author} {\bibfnamefont {M.}~\bibnamefont {Berz}},
  \bibinfo {author} {\bibfnamefont {E.}~\bibnamefont {Valetov}}, \bibinfo
  {author} {\bibfnamefont {S.}~\bibnamefont {Chekmenev}}, \bibinfo {author}
  {\bibfnamefont {S.}~\bibnamefont {Andrianov}}, \ and\ \bibinfo {author}
  {\bibfnamefont {A.}~\bibnamefont {Ivanov}} (\bibinfo {collaboration} {on
  behalf of the JEDI Collaboration}),\ }in\ \href {\doibase
  doi:10.18429/JACoW-ICAP2015-MODBC4} {\emph {\bibinfo {booktitle} {Proceedings
  of ICAP 2015, Shanghai, China}}}\ (\bibinfo {year} {2015})\ \bibinfo {note}
  {{MODBC4}}\BibitemShut {NoStop}%
\bibitem [{\citenamefont {Makino}\ and\ \citenamefont {Berz}(2006)}]{COSY}%
  \BibitemOpen
  \bibfield  {author} {\bibinfo {author} {\bibfnamefont {K.}~\bibnamefont
  {Makino}}\ and\ \bibinfo {author} {\bibfnamefont {M.}~\bibnamefont {Berz}},\
  }\href {\doibase 10.1016/j.nima.2005.11.109} {\bibfield  {journal} {\bibinfo
  {journal} {Nuclear Instruments and Methods in Physics Research, Section A:
  Accelerators, Spectrometers, Detectors and Associated Equipment}\ }\textbf
  {\bibinfo {volume} {558}},\ \bibinfo {pages} {346} (\bibinfo {year}
  {2006})}\BibitemShut {NoStop}%
\bibitem [{\citenamefont {Berz}(1999)}]{ModernMapMethods}%
  \BibitemOpen
  \bibfield  {author} {\bibinfo {author} {\bibfnamefont {M.}~\bibnamefont
  {Berz}},\ }\href@noop {} {\emph {\bibinfo {title} {Modern map methods in
  particle beam physics}}},\ Advances in Imaging and Electron Physics\
  (\bibinfo  {publisher} {Academic Press},\ \bibinfo {address} {San Diego,
  CA},\ \bibinfo {year} {1999})\BibitemShut {NoStop}%
\bibitem [{\citenamefont {Valetov}\ \emph {et~al.}(2015)\citenamefont
  {Valetov}, \citenamefont {Berz},\ and\ \citenamefont
  {Senichev}}]{Valetov2015}%
  \BibitemOpen
  \bibfield  {author} {\bibinfo {author} {\bibfnamefont {E.}~\bibnamefont
  {Valetov}}, \bibinfo {author} {\bibfnamefont {M.}~\bibnamefont {Berz}}, \
  and\ \bibinfo {author} {\bibfnamefont {Y.}~\bibnamefont {Senichev}} (\bibinfo
  {collaboration} {on behalf of the JEDI Collaboration}),\ }in\ \href {\doibase
  10.18429/JACoW-ICAP2015-THDBC2} {\emph {\bibinfo {booktitle} {Proceedings of
  ICAP 2015, Shanghai, China}}}\ (\bibinfo {year} {2015})\ \bibinfo {note}
  {{THDBC2}}\BibitemShut {NoStop}%
\bibitem [{Note1()}]{Note1}%
  \BibitemOpen
  \bibinfo {note} {In sextupole family names, 'F' denotes focusing, 'D'
  defocusing, 'P' positive dispersion, and 'N' negative dispersion. The
  \protect \emph {BNL} lattice only has positive dispersion sextupole
  families.}\BibitemShut {Stop}%
\bibitem [{\citenamefont {Senichev}\ \emph {et~al.}(2016)\citenamefont
  {Senichev}, \citenamefont {Andrianov}, \citenamefont {Berz}, \citenamefont
  {Chekmenev}, \citenamefont {Ivanov}, \citenamefont {Lorentz}, \citenamefont
  {Pretz},\ and\ \citenamefont {Valetov}}]{Senichev2016}%
  \BibitemOpen
  \bibfield  {author} {\bibinfo {author} {\bibfnamefont {Y.}~\bibnamefont
  {Senichev}}, \bibinfo {author} {\bibfnamefont {S.}~\bibnamefont {Andrianov}},
  \bibinfo {author} {\bibfnamefont {M.}~\bibnamefont {Berz}}, \bibinfo {author}
  {\bibfnamefont {S.}~\bibnamefont {Chekmenev}}, \bibinfo {author}
  {\bibfnamefont {A.}~\bibnamefont {Ivanov}}, \bibinfo {author} {\bibfnamefont
  {B.}~\bibnamefont {Lorentz}}, \bibinfo {author} {\bibfnamefont
  {J.}~\bibnamefont {Pretz}}, \ and\ \bibinfo {author} {\bibfnamefont
  {E.}~\bibnamefont {Valetov}} (\bibinfo {collaboration} {on behalf of the JEDI
  Collaboration}),\ }in\ \href {\doibase 10.18429/JACoW-IPAC2016-THPMR005}
  {\emph {\bibinfo {booktitle} {Proceedings of IPAC 2016, Busan, Korea}}}\
  (\bibinfo {year} {2016})\ \bibinfo {note} {{THPMR005}}\BibitemShut {NoStop}%
\end{thebibliography}%

\end{document}